\documentclass[11pt,english]{article}
\usepackage[latin9]{inputenc}
\usepackage{color}
\usepackage{array}
\usepackage{multirow}
\usepackage{amsmath}
\usepackage{graphicx}
\usepackage{setspace}
\numberwithin{equation}{section}

\usepackage{esint}
\usepackage{latexsym}
\usepackage{graphicx}\usepackage{bm}\usepackage{longtable}

\def\l{{\lambda}}

\def\d{{\delta}}

\def\o{{\omega}}
\def\O{{\Omega}}
\def\e{{\epsilon}}

\def\a{{\alpha}}
\def\b{{\beta}}
\def\c{{\chi}}

\def\g{{\gamma}}
\def\G{{\Gamma}}

\def\h{\eta}
\def\p{{\pi}}

\def\m{{\mu}}
\def\n{{\nu}}
\def\r{{\rho}}
\def\s{{\sigma}}

\def\th{{\theta}}

\def\ps{{\psi}}

\def\Ps{{\Psi}}

\def\({\left(}
\def\){\right)}
\def\[{\left[}
\def\]{\right]}

\newcommand{\pd}{{\partial}}
\newcommand{\dg}{\dagger}
\newcommand{\pr}{\parallel}
\newcommand{\pp}{\perp}

\newcommand{\Tr}{\text{Tr}}

\usepackage{xcolor}

\date{}

\usepackage[english]{babel}
\begin{document}
	\title{\bf Spin polarization of holographic baryon in strongly coupled fluid}
\maketitle
\vspace{-12mm}
\begin{center}
Si-wen Li$^{*}$\footnote{siwenli@dlmu.edu.cn (co-corresponding author)}, 
Shu Lin$^{\dagger}$\footnote{linshu8@mail.sysu.edu.cn},
\par\end{center}

\begin{center}
\emph{$^{1}$School of Science, Dalian Maritime University, Dalian
116026, China}
\par\end{center}

\begin{center}
$^{2}$\textsl{School of Physics and Astronomy, Sun Yat-Sen University,
Zhuhai 519082, China}
\par\end{center}


\vspace{6mm}

\begin{abstract}

The spin polarization for baryon in a hydrodynamic medium has been extensively studied in the weakly coupled regime using quantum kinetic theory. As a first study of this problem in the strongly coupled regime, we investigate holographically the spectral function of a probe baryon in the fluid-gravity background. This is done by carefully performing gradient expansion of the Dirac equation in the fluid-gravity background. Different contributions in the expansion are understood in terms of density matrices of the probe baryon and the medium. The resulting spectral functions indicate that the holographic baryonic is polarized as responses to fluid acceleration, shear stress and vorticity. The structures of the responses are similar to those found in weakly coupled studies.

\newpage{}
\end{abstract}

\section{Introduction}

Spin polarization is one of the key observables in particle physics,
which is a probe on the interaction among particles. Recently there
has been growing interest in spin polarization in relativistic heavy
ion collisions: based on early idea of spin-orbit coupling, baryons
and vector mesons are predicted to have spin polarization and spin
alignment respectively \cite{Liang:2004ph,Liang:2004xn}. These have indeed been confirmed
by heavy ion experiments \cite{STAR:2017ckg}. A distinguishing feature of heavy
ion experiments is the presence of rapid spinning quark-gluon plasma
(QGP), which encodes the collective motion of quarks and gluons and
adopts hydrodynamic description in terms of vorticity. The measurements
of global polarization of $\Lambda$ hyperon have been well understood
with spin-vorticity coupling \cite{Becattini:2013fla,Fang:2016vpj,Li:2017slc,Wei:2018zfb,Liu:2019krs,Becattini:2017gcx}. More recently, it has been
realized that spin also coupled to other hydrodynamic gradients \cite{Hidaka:2017auj,Liu:2021uhn,Becattini:2021suc},
which is crucial for understanding measurements of local polarization
\cite{Fu:2021pok,Becattini:2021iol,Yi:2021ryh}.

A notable framework in understanding spin polarization is the quantum kinetic
theory (QKT), which is based on quasi-particles with spin degrees of freedom and incorporates collisional effect systematically, see \cite{Hidaka:2022dmn} for a recent review. Useful insights have been gained by studies within QKT framework: we can separate the spin polarization into a local equilibrium contribution and a steady state contribution. The first one is induced through spin response to hydrodynamic gradients. The free theory limit of this contribution is widely known \cite{Becattini:2013fla,Becattini:2021suc,Liu:2021uhn} and extensively implemented in phenomenological studies \cite{Fu:2021pok,Becattini:2021iol,Yi:2021ryh}. The second one is unique to interacting theories. In the presence of dissipative gradient, such as shear and temperature gradient but not vorticity, the fermions will reach a steady state, in which the nonvanishing collision with other particles in the medium balances the effective force exerted by the dissipative gradient. The deviation of steady state from local equilibrium state, as well as the resulting contribution to spin polarization depend on details of interaction \cite{Lin:2022tma,Lin:2024zik,Wang:2024lis,Fang:2024vds}. 
In many phenomenology motivated theoretical studies, a strange quark is often considered as a probe to off-equilibrium QGP. This allows us to separate the steady state effect of the medium and the probe \cite{Lin:2022tma,Wang:2024lis,Fang:2024vds}. More recently, general interaction effect has been classified into modifications to spectral function, Kubo-Martin-Schwinger relation and distribution function \cite{Lin:2024svh}. Radiative correction to spectral function has been considered in \cite{Fang:2023bbw,Lin:2024svh}

Despite success of QKT framework, the effectiveness of quasi-particle picture in QGP should not be taken for granted: phenomenological studies of QGP point to a low viscosity, which is close to the viscosity lower bound \cite{Song:2010mg}. Such a low viscosity suggests QGP might be strongly coupled, for which quasi-particles might not exist. The purpose of this work is to provide a study of spin polarization in the regime of strong coupling using gauge-gravity duality \cite{Maldacena:1997re,Aharony:1999ti,Witten:1998qj}. While there is no quasi-particle in strongly coupled regime, we will see it is still possible to understand contributions from different origins in terms of density matrix, which can be further separated into density matrices of the probe and medium in the probe limit. We will find the spectral function of a holographic baryon in the fluid-gravity background. The correction to the spectral function in the background encodes the spin polarization of the baryon in response to hydrodynamic gradients. We will see the corresponding spectral function splits naturally into local equilibrium state contribution and steady state one. A similar study has been performed in the weakly coupled case, but for local equilibrium state only \cite{Lin:2024svh}. This work also provides a first example of steady state contribution to spectral function\footnote{The steady state effect in the QKT framework corrects the spin polarization, but not the spectral function.}.

The paper is structured as follows: we first present in Sec.~\ref{sec:general}
a general analysis on the structure of correlation function and constraint from parity; we then discuss how to perform consistent gradient expansion in fluid gravity background in Sec.~\ref{sec:holography}; Complete calculations of the retarded and spectral functions are presented in Sec.~\ref{sec:retarded}. The spectral function is found to be spin polarized in responses to fluid acceleration, shear stress and vorticity; Conclusion
and outlook are devoted to Sec.~\ref{sec:conclusion}. Notations and some explicit expressions are collected in two appendices.

\section{General structure of fermionic correlation function}\label{sec:general}

Since the holographic baryon we study consists of two Weyl fermions of opposite helicities, we first discuss correlation functions of Weyl fermion with one helicity. The retarded and advanced fermionic correlation function are defined respectively as,
\begin{align}\label{G_RA}
&G^R_{\alpha\beta}\left(X,p\right)=i\int d^{4}(x-y)e^{ip\cdot (x-y)}\theta\left(x^0-y^0\right)\left\langle \Psi_{\alpha}\left(x\right)\Psi_{\beta}^{\dagger}\left(y\right)+\Psi_{\beta}^\dg\left(y\right)\Psi_{\alpha}\left(x\right)\right\rangle ,\nonumber\\
&G^A_{\alpha\beta}\left(X,p\right)=-i\int d^{4}(x-y)e^{ip\cdot( x-y)}\theta\left(-x^0+y^0\right)\left\langle \Psi_{\alpha}\left(x\right)\Psi_{\beta}^{\dagger}\left(y\right)+\Psi_{\beta}^\dg\left(y\right)\Psi_{\alpha}\left(x\right)\right\rangle,
\end{align}
with $\Psi$ being a composite gauge invariant Weyl operator. $X=\frac{x+y}{2}$ and $p^\m=(\o,\vec{p})$. The fermionic correlator is to be evaluated in a fluid background characterized by gradients of temperature and fluid velocity (to be denoted by $\pd_X$). We require $\o,\,p\gg\pd_X$ such that we can interpret $X$ and $p$ as the coarse-grained position and momentum of the baryon. $\alpha,\beta$ denote Weyl indices. $G_{R}$ is a $2\times2$ matrix. The expectation values are to be understood as $\Tr[D\cdots]$ with $D$ being the density matrix. We will consider a probe baryon in a fluid background, in which the density matrix can be factorized as $D=D_{\text{probe}}\otimes D_{\text{fluid}}$. Since $D$ is hermitian, we can easily show
\begin{align}\label{GRA}
G^{R}(\o,\vec{p})^\dg=G^A(\o,\vec{p}).
\end{align}
A closely related quantity is the spectral function defined similarly as,
\begin{equation}\label{rho}
\rho_{\alpha\beta}\left(X=\frac{x+y}{2},p\right)=\int d^{4}(x-y)e^{ip\cdot (x-y)}\left\langle \Psi_{\alpha}\left(x\right)\Psi_{\beta}^{\dagger}\left(y\right)+\Psi_{\beta}^\dg\left(y\right)\Psi_{\alpha}\left(x\right)\right\rangle .
\end{equation}
By definition, we have
\begin{align}\label{rho_GR}
i\r(X,p)=G^R(X,p)-G^A(X,p)=G^R(X,p)-G^R(X,p)^\dg,
\end{align}
where we have used \eqref{GRA} in the last equality. It follows that $\r=\r^\dg$.

To model a baryon field, we shall combine the contribution from both helicities. We shall perform explicit calculations for one helicity only. The contribution from the other helicity will be obtained by parity transformation. We now derive the precise relation. We start with the following identity
\begin{align}\label{parity}
\langle \O_b|\p^\dg\p O\p^\dg\p|\O_a\rangle=\langle \widetilde{\O}_b|\p O\p^\dg|\widetilde{\O}_a\rangle,
\end{align}
where $\p$ is the parity operator and $|\widetilde{\O}_{a/b}\rangle=\p|\O_{a/b}\rangle$. The matrix element is readily generalizable to expectation value, in which the in and out states are replaced by trace over density matrix. To be specific, let us take the Weyl fermion to be left-handed, and denote the corresponding density matrix to be $D_{\text{L}}\otimes D_{\text{fluid}}$. Parity transformation turns the fermion to be right-handed and also acts on the fluid background, with the transformed density matrix denoted by $D_{\text{R}}\otimes \widetilde{D}_{\text{fluid}}$. We take the operator to be
\begin{align}\label{operator}
O_{\a\b}(x;y)=\Ps_\a^L(x)\Ps_\b^{L\dg}(y)+\Ps_\b^{L\dg}(y)\Ps_\a^L(x),
\end{align}
where we have used superscript $L$ to indicate the Weyl field being left-handed. The parity transformation of $O_{\a\b}$ is given by
\begin{align}
\p O_{\a\b}(x^0,\vec{x};y^0,\vec{y})\p^\dg=\Ps_\a^R(x^0,-\vec{x})\Ps_\b^{R\dg}(y^0,-\vec{y})+\Ps_\b^{R\dg}(y^0,\vec{y})\Ps_\a^R(x^0,-\vec{x}).
\end{align}
The net effect is the following: the operator remains the same but with the field replaced by the right-handed counterpart. The argument is spatially inverted. Applying Wigner transform, we easily find
\begin{align}\label{rho_trs}
\r_{\a\b}^L(X,\o,\vec{p})=\widetilde{\r}_{\a\b}^R(X,\o,-\vec{p}).
\end{align}
Here the tilde indicates the parity transformation on the fluid background. Similar relation can be derived for retarded and advanced correlators, which we do not elaborate.

\section{Holographic prescription and transformations}\label{sec:holography}

In heavy ion collisions, the baryons form at freezeout as end point of QGP evolution. The hydrodynamic gradients at freezeout leave imprints on spectral function of baryons in the form of gradient corrections, contributing a part of the spin polarization of baryons. Motivated by this picture, we consider holographic baryons in QGP fluid. The former is modeled by bulk Dirac fermions in a bottom-up approach, see also \cite{Nakas:2020hyo, Li:2023wyb} for top-down approaches. The latter is modeled by the fluid-gravity background.
The holographic prescription for retarded function of fermionic operators has been established in \cite{Henningson:1998cd,Mueck:1998iz,Henneaux:1998ch}. We will follow the notation of \cite{Iqbal:2009fd} below. However, the formulation in Fourier component assumes translational invariance, which does not hold in fluid background we use. We shall generalize the prescription to coordinate space. For pedagogical purpose, we start with Dirac action in AdS-Schwarzschild background
\begin{align}
ds^2=\frac{-f(z)dt^2+d\vec{x}^2+dz^2/f(z)}{z^2},
\end{align}
with $f(z)=1-\frac{z^4}{b^4}$.
By using integration by part and the bulk Dirac equation $(\G^M\nabla_{M}-m)\ps=0$, we obtain the boundary term as
\begin{align}\label{action}
	S&=\int d^{5}x \sqrt{-g} \overline{\psi}(x)i\left( \Gamma^{M} \nabla_{M}-m\right) \psi(x) \notag \\
	&=i\int d^{4}x \sqrt{-g} g_{zz}^{-1 / 2} \overline{\psi}(x) \gamma^{\hat{z}} \psi(x)\big|_{z \to 0}+...
\end{align}
where $\G^z=e^z_{\hat{z}}\g^{\hat{z}}=g_{zz}^{-1/2}\g^{\hat{z}}$. We shall not need explicit form of the Dirac equation and representation of gamma matrices, which are collected in appendix~\ref{sec:appA}. For now, we need
\begin{align}\label{gammas}
\g^{\hat{z}}
=\begin{pmatrix}
	I& 0\\
	0& -I
\end{pmatrix},\quad
\ps=\begin{pmatrix}
\ps_R\\
\ps_L
\end{pmatrix}\equiv(-g g^{zz})^{-1/4}
\begin{pmatrix}
\c_R\\
\c_L
\end{pmatrix}.
\end{align}
Asymptotic analysis gives \cite{Iqbal:2009fd}
\begin{align}\label{asymptotic}
\c_L=A z^{-m}+B z^{1+m},\quad \c_R=C z^{1-m}+D z^m.
\end{align}
For $m\ge0$($m\le0$), $\c_L$($\c_R$) should be chosen as source to fermionic operator $\Ps$. We shall use two Weyl fermions with opposite helicity to make a holographic baryon. For this reason, we choose $m=0$. Plugging \eqref{gammas} into \eqref{action}, we obtain
\begin{align}
S=\int d^4x\(-\c_L^{b\dg}(x)\c_R^b(x)+\c_R^{b\dg}\c_L^b(x)\).
\end{align}
with $\c_{R/L}^{b}=\c_{R/L}(z\to0)$. To be specific we choose $\c_L^b$ as the source in the actual calculation, corresponding to correlator of left-handed Weyl operator. The counterpart for right-handed Weyl fermion will be deduced by \eqref{rho_trs}. Imposing the infalling boundary condition at the horizon, we have the following retarded correlator
\begin{align}\label{GR_def}
G_R(x,y)=-\frac{\d^2S}{\d\c_L^{b\dg}(x)\d\c_L^b(y)}=\frac{\d \c_R^b(x)}{\d \c_L^b(y)}.
\end{align}
The prescription will be generalized to the fluid gravity background, which is given by \cite{Bhattacharyya:2007vjd}
\begin{align}\label{fluid_background}
ds^{2}=& -2u_{\mu}dx^{\mu}dr - r^{2}f(br)u_{\mu}u_{\nu}dx^{\mu}dx^{\nu}+r^{2}P_{\mu \nu}dx^{\mu}dx^{\nu}\nonumber\\
& + 2r^{2}bF(br)\sigma_{\mu \nu}dx^{\mu}dx^{\nu}+\frac{2}{3}ru_{\mu}u_{\nu}\partial_{\lambda}u^{\lambda}dx^{\mu}dx^{\nu}-ru^{\lambda}\partial_{\lambda}(u_{\nu}u_{\mu})dx^{\mu}dx^{\nu},
\end{align}
with
\begin{align}
F(r)=\int_{r}^{\infty} dx\frac{x^{2}+x + 1}{x(x + 1)(x^{2}+1)}=\frac{1}{4}\left[\ln\left(\frac{(1 + r)^{2}(1 + r^{2})}{r^{4}}\right)-2\arctan(r)+\pi\right].
\end{align}
The first line of \eqref{fluid_background} is simply a boosted Schwarzschild background written in Eddington-Finkelstein (EF) coordinates with $u_\m(x)$ and $b(x)$ being slow-varying velocity and inverse temperature. The explicit relations are given by
\begin{align}
r=\frac{1}{z},\quad dt=d\bar{t}-\frac{dz}{f(z)},\quad dx^i=d\bar{x}^i,
\end{align}
with the barred coordinate corresponding to Schwarzschild coordinate.
The second line of \eqref{fluid_background} is the gradient correction to the boosted metric, with the three terms corresponding to shear, bulk stress and acceleration in local rest frame of fluid respectively. Explicit form of the metric in local rest frame of the fluid is collected in appendix~\ref{sec_appB} for completeness. The two lines have clear physical interpretation: the first line corresponds to local equilibrium state of QGP, while the second line characterizes steady state deviation from the local equilibrium one. They encode local equilibrium part and gradient correction part of $D_\text{fluid}$. Since the bulk stress carries no spatial index, it cannot couple to spin polarization. We shall not consider bulk stress in this paper.

Before making the generalization, it is useful to discuss symmetry transformations of the correlator. The bulk action is invariant under two symmetries: local Lorentz transformation (LLT) and diffeomorphism by construction, but the boundary correlator is not invariant. Under the LLT, the bulk fermion transforms as
\begin{align}
\ps_L(x)\to M_L(x)\ps_L(x),\quad \ps_L^\dg(y)\to \ps_L^\dg(y)M_R(y)^{-1},
\end{align}
where $M_L(x)$ and $M_R(y)$ are Lorentz transformations on 5D fermion. On the boundary, the LLT reduces to
\begin{align}
\c_L^b(x)\to M_L^b(x)\c_L^b(x),\quad \c_L^{b\dg}(y)\to \c_L^{b\dg}(y)M_R^b(y)^{-1},
\end{align}
with $M_L^b=M_L(z\to0)$ and $M_R^b=M_R(z\to0)$ in the above. It follows that
\begin{align}
G_R(x,y)\to M_R^b(x)^{-1}G_R(x,y)M_L^b(y).
\end{align}
In order to fix the ambiguity, we need to choose a frame. A natural choice is to require the boundary fermion to be in the same frame as the bulk fermion. This is most clearly written for AdS-Schwarzschild background as
\begin{align}\label{vielbein_schw}
\bar{e}_t^{\hat{t}}\to\frac{1}{z},\quad \bar{e}_i^{\hat{i}}\to\frac{1}{z},\quad \bar{e}_z^{\hat{z}}\to\frac{1}{z},
\end{align}
as $z\to0$. We have used barred vielbein to indicate they correspond to Schwarzschild coordinate. Hatted and unhatted coordinates correspond to flat and curved coordinates respectively. Note that $M_{L/R}^b$ include both 4D Lorentz transformation and special conformal transformation. The choice \eqref{vielbein_schw} eliminates both freedoms. The frame choice is readily generalized to the fluid gravity background. We first note that the background at $O(\pd^0)$ is nothing but boosted AdS-Schwarzschild metric. Note that it corresponds to a 4D boost on the boundary, which brings a static fluid element to a moving fluid element. If we choose local rest frame, the local metric is simple AdS-Schwarzschild, then the appropriate frame choice is obtained by transforming \eqref{vielbein_schw} to EF coordinates using $e_M^a=\bar{e}_N^a\frac{\pd \bar{x}^N}{\pd x^M}$. We obtain
\begin{align}\label{vielbein_EF}
e_t^{\hat{t}}\to\frac{1}{z},\quad e_t^{\hat{z}}\to\frac{1}{z},\quad e_i^{\hat{i}}\to\frac{1}{z},\quad e_z^{\hat{z}}\to\frac{1}{z}.
\end{align}
Furthermore, the gradient correction to the boost metric is subleading as $z\to0$, which means the asymptotic behavior \eqref{vielbein_EF} fixing the frame choice is unaffected by the subleading correction.

Turning to diffeomorphism, we note that the bulk spinor is invariant under diffeomorphism. Although the definitions of $\c_{R/L}$ in \eqref{gammas} do seem to introduce dependence on choice of coordinate through the factor $(-g g^{zz})^{-1/4}$, the factor is introduced merely to remove extra powers of $z$ at the boundary. The powers can be related to scaling dimension of boundary operator that are uniquely determined from asymptotic AdS metric. We may insist using the same numerical factor in defining $\c_{R/L}$ in fluid gravity background. This way the bulk diffeomorphism leaves no freedom on the boundary correlator\footnote{In principle, the factor we use instead of powers of $z$ could bring in possible contact terms to the boundary correlator. This does not occur for the choice $m=0$.}.

\section{Gradient expansion and retarded correlator}\label{sec:retarded}

In this section, we study first order gradient correction to the retarded correlator, which arises from the following gradient correction to density matrix $D_\text{fluid}^{(0)}\otimes D_\text{baryon}^{(1)}$ and $D_\text{fluid}^{(1)}\otimes D_\text{baryon}^{(0)}$, with superscripts denoting orders of gradient. The gradient expansion of the Dirac equation involves expansions of the Dirac field and the Dirac operator separately. which are to be identified with $D_\text{fluid}^{(0)}\otimes D_\text{baryon}^{(1)}$ and $D_\text{fluid}^{(1)}\otimes D_\text{baryon}^{(0)}$ respectively. We shall first perform expansion of the Dirac field in the boosted AdS-Schwarzschild background, then expansion of the Dirac operator.

\subsection{Gradient correction to the Dirac field}
In a translational background like AdS-Schwarzschild, we can apply Fourier transform to \eqref{GR_def} to obtain
\begin{align}
\c_R^b(p)=G_R(p)\c_L^b(p).
\end{align}
This allows us to solve the bulk EOM for $\c_{R/L}(z,p)$ and extract $G_R(p)$ from the boundary behavior. In fluid gravity background, $G_R(x,y)$ is to be Wigner transformed as in \eqref{GRA} and to be calculated using gradient expansion. The tricky point is that we can do gradient expansion for $G_R(x,y)$, but not $\c_{R}(x)$ and $\c_L(y)$ separately, as the gradient corresponds to variation with respect to $X$. We instead perform gradient expansion to the bulk extension of $G_R(x,y)$, which is the boundary-bulk propagator $G_R(x,z|y)$ satisfying
\begin{align}\label{bb_prop}
\c_R(x,z)=\int_y G(x,z|y)\c_L^b(y),
\end{align}
with $\int_y=\int d^4y$. Note that $G(x,z|y)$ satisfies the same EOM as $\c_R(x,z)$. We proceed with the boosted AdS-Schwarzschild background in EF coordinate and choose the following vielbein
\begin{align}\label{vielbein_local}
e_t^{\hat{t}}=\frac{f^{1/2}}{z},\;e_t^{\hat{z}}=\frac{1}{z f^{1/2}},\; e_i^{\hat{i}}=\frac{1}{z},\;e_z^{\hat{z}}=\frac{1}{ f^{1/2}}.
\end{align}
All other components vanish. The resulting Dirac equation in terms of components $\c_{R/L}$ reads
\begin{align}\label{EOM_RL}
f^{1 / 2} \(\partial_{z}-\frac{\pd_0^x}{f}\) \chi_{R}(x,z)+i D_{\mu}^{x} \sigma^{\mu} \chi_{L}(x,z)&=0, \nonumber\\
f^{1 / 2} \(\partial_{z}-\frac{\pd_0^x}{f}\) \chi_{L}(x,z)-i D_{\mu}^{x} \overline{\sigma}^{\mu} \chi_{R}(x,z)&=0,
\end{align}
with $D_{\mu}^{x}=\left(f^{-1 / 2} \partial_{0}^{x}, \partial_{i}^{x}\right)$. We can derive a second order EOM for $\c_R$ from the two first order equations
\begin{equation}\label{EOM_R2}
\c_R''+\frac{f'}{2f}\c_R'-\frac{2D_0^x}{f^{1/2}}\c_R'+\frac{D_0^xf'}{2f^{3/2}}\c_R+\frac{\pd_i^x\pd_i^x}{f}\c_R-\frac{iD_0^xf'}{2f^{3/2}}\c_L=0.
\end{equation}
We have suppressed the common argument $(x,z)$ and denoted $\pd_z$ by prime.
To express $\c_L$ in the last term in terms of $\c_R$, we act on the first line of \eqref{EOM_RL} by $D_\n^x\bar{\s}^\n$ to obtain
\begin{align}
D^{x}\cdot\overline{\sigma}\(f^{1/2}\chi_{R}'-D_0^x\c_R\)-iD_x^{2}\chi_{L}&=0.
\end{align}
We can solve for $\c_L$ by formally dividing $D_x^2$ and plug into \eqref{EOM_R2} to arrive at
\begin{align}\label{EOM_R}
\c_R''+\frac{f'}{2f}\c_R'-\frac{2D_0^x}{f^{1/2}}\c_R'+\frac{D_0^xf'}{2f^{3/2}}\c_R+\frac{\pd_i^x\pd_i^x}{f}\c_R+\frac{D_0^xf'}{2f^{3/2}}\frac{D^x\cdot\bar{\s}f'(f^{1/2}\chi_{R}'-D_0^x\c_R)}{-D_x^2}=0.
\end{align}
The same equation is satisfied by $G(x,z|y)$. The Wigner transform of $G(x,z|y)$ is defined as
\begin{align}\label{Wigner_G}
G(X,p,z)=\int_s e^{-ip\cdot s}G(x,z|y),
\end{align}
with $s=x-y$. We can then perform gradient expansion using
\begin{align}
&\pd^x_\m=\pd^s_\m+\frac{1}{2}\pd^X_\m\to ip_\m+\frac{1}{2}\pd^X_\m.
\end{align}
Denoting $P^\m=(f^{-1/2}\o,p_i)$ and $D_\m^X=(f^{-1/2}\pd_0^X,\pd_i^X)$, we easily find the following EOM expanded up to $O(\pd^X)$
\begin{align}\label{EOM_G}
&G''+2i\frac{\omega}{f}G'+\frac{f'}{2f}G'-i\omega\frac{f'}{2f^{2}}G-\frac{p^{2}}{f}G+\frac{f'}{2f^{3/2}}\omega \frac{P\cdot\overline{\sigma}}{P^{2}}\left(G'+\frac{i\omega}{f}G\right)+\bigg[-\frac{\pd_0^X}{f}G'+\frac{f'}{4f^2}\pd_0^XG\nonumber\\
&+ip_i\frac{\pd_i^X}{f}G+\frac{f'}{4f^{3/2}}\frac{P\cdot\bar{\s}}{P^2}i\pd_0^X\(G'+\frac{2i\o}{f}G\)+\frac{f'}{2f^{3/2}}i\o\(\frac{D^X\cdot\bar{\s}}{-2P^2}+\frac{P\cdot D_X P\cdot\bar{\s}}{(P^2)^2}\)\(G'+\frac{i\o}{f}G\)\bigg]\nonumber\\
&=0.
\end{align}
In arriving at the equation above, we assume the operator $D_x^2$ is invertible. It would be violated if $P^2=-\o^2/f(z)+p^2$ vanish in the bulk. This does not occur for time like momenta, which is what we need for baryonic excitation. It is possible to consider excitation with spacelike momenta by introducing $i\e$ prescription. We shall restrict ourselves to timelike momenta below. 
The boundary condition can be deduced from the first line of \eqref{EOM_RL} and \eqref{bb_prop} as
\begin{align}
-iD_\m^x\s^\m\c_L(x,z)=\(f^{1/2}\pd_z\c_R(x,z)-D_0^x\c_R\)=\int_y\(f^{1/2}\pd_z-D_0^x\)G(x,z|y)\c_L^b(y).
\end{align}
In the limit $z\to0$, we have
\begin{align}
\(\pd_z-\pd_0^x\)G(x,z|y)\to-i\pd^x\cdot\s\d(x-y).
\end{align}
Its Wigner transform gives the boundary condition for $G(X,p,z)$
\begin{align}\label{bc}
\(\pd_z+i\o-\frac{1}{2}\pd_0^X\)G(X,p,z\to0)\to p\cdot\s.
\end{align}
The boundary correlator is given by $G_R(x,y)=G(x,z\to0|y)$, or its Wigner transformed version
\begin{align}
G_R(X,p)=G_R(X,p,z\to0).
\end{align}

\subsection{Retarded correlator from gradient correction to the probe}\label{sec:probe}

\begin{figure}
	\centering
	\includegraphics[width=0.5\linewidth]{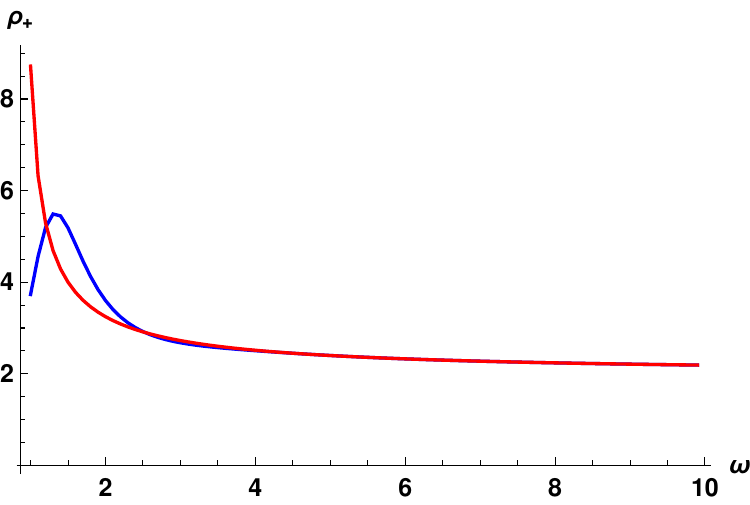}
	\caption{$\r_+$ versus $\o$ for $p=0.9$ plotted in blue. The unit is set by $\p T=1$. We also plot the counterpart from vacuum spectral function in orange, whose analytic form is known \cite{Iqbal:2009fd}: $\r_+=2\text{sgn}(\o)\th(\o^2-p^2)\(\frac{\o+p}{\o-p}\)^{1/2}$. The effect of the medium softens the singularity in vacuum spectral function.}
	\label{fig:G0}
\end{figure}

We now solve \eqref{EOM_G} in gradient expansion as
\begin{align}
G=G_{0}+G_1+\cdots,
\end{align}
with subscript indicating order of gradient. $G_0$ satisfies
\begin{align}\label{EOM_G0}
&G_0''+2i\frac{\omega}{f}G_0'+\frac{f'}{2f}G_0'-i\omega\frac{f'}{2f^{2}}G_0-\frac{p^{2}}{f}G_0+\frac{f'}{2f^{3/2}}\omega \frac{P\cdot\overline{\sigma}}{P^{2}}\left(G_0'+\frac{i\omega}{f}G_0\right)=0.
\end{align}
By rotational invariance, we can decompose $G_0=A(\o,p)+B(\o,p)\hat{p}\cdot\s$. Plugging this into \eqref{EOM_G0}, we obtain the following coupled EOM from the coefficients of $1$ and $\hat{p}\cdot\s$
\begin{align}\label{EOM_AB}
&-\frac{p^{2}A}{f}+\frac{2i\omega A'}{f}-\frac{i\omega^{3}A f'}{2\left(p^{2}-\frac{\omega^{2}}{f}\right)f^{3}}+\frac{ip\omega^{2}B f'}{2\left(p^{2}-\frac{\omega^{2}}{f}\right)f^{5/2}}-\frac{i\omega A f'}{2f^{2}}-\frac{\omega^{2}A' f'}{2\left(p^{2}-\frac{\omega^{2}}{f}\right)f^{2}}+\frac{A' f'}{2f}\nonumber\\
&+\frac{p\omega B' f'}{2\left(p^{2}-\frac{\omega^{2}}{f}\right)f^{3/2}}+A'' = 0\\
&-\frac{p^{2}B}{f}+\frac{2i\omega B'}{f}-\frac{i\omega^{3}B f'}{2\left(p^{2}-\frac{\omega^{2}}{f}\right)f^{3}}+\frac{ip\omega^{2}A f'}{2\left(p^{2}-\frac{\omega^{2}}{f}\right)f^{5/2}}-\frac{i\omega B f'}{2f^{2}}-\frac{\omega^{2}B' f'}{2\left(p^{2}-\frac{\omega^{2}}{f}\right)f^{2}}+\frac{B' f'}{2f}\nonumber\\
&+\frac{p\omega B' f'}{2\left(p^{2}-\frac{\omega^{2}}{f}\right)f^{3/2}}+B'' = 0.
\end{align}
We can easily decouple the EOM with the linear combination $A_\pm(\o,p)=A(\o,p)\pm B(\o,p)$, which satisfy
\begin{align}\label{EOM_Apm}
&-\frac{p^{2}A_\pm}{f}+\frac{2i\omega A_\pm'}{f}-\frac{i\omega^{3}A_\pm f'}{2\left(p^{2}-\frac{\omega^{2}}{f}\right)f^{3}}\pm\frac{ip\omega^{2}A_\pm f'}{2\left(p^{2}-\frac{\omega^{2}}{f}\right)f^{5/2}}-\frac{i\omega A_\pm f'}{2f^{2}}-\frac{\omega^{2}A_\pm' f'}{2\left(p^{2}-\frac{\omega^{2}}{f}\right)f^{2}}+\frac{A_\pm' f'}{2f}\nonumber\\
&\pm\frac{p\omega A_\pm' f'}{2\left(p^{2}-\frac{\omega^{2}}{f}\right)f^{3/2}}+A_\pm'' = 0.
\end{align}
The boundary condition from \eqref{bc} reduces at $O(\pd^0)$ to
\begin{align}
(\pd_z+i\o)A_\pm(z\to0)\to-\o\mp p.
\end{align}
The other boundary condition at the horizon is regularity condition for $A_\pm$. Without actually solving the equation, we can easily deduce the symmetry property $A_+(\o,p,z)=A_-(\o,-p,z)$. Note that the boundary correlator at $O(\pd^0)$ is given by
\begin{align}
G_R^{(0)}(\o,p)=A_+(\o,p,z\to0)\frac{1+\hat{p}\cdot\s}{2}+A_-(\o,p,z\to0)\frac{1-\hat{p}\cdot\s}{2}.
\end{align}
\eqref{EOM_Apm} is solved numerically and the spectral function is extracted as
\begin{align}
\r^L=\text{Im}[A_+(\o,p,z\to0)]\frac{1+\hat{p}\cdot\s}{2}+\text{Im}[A_-(\o,p,z\to0)]\frac{1-\hat{p}\cdot\s}{2}\equiv \r_+\frac{1+\hat{p}\cdot\s}{2}+\r_-\frac{1-\hat{p}\cdot\s}{2}.
\end{align}
We have reinstated the superscript $L$ for left-handed fermionic operator. The structures $\frac{1\pm\hat{p}\cdot\vec{\s}}{2}$ have positive and negative helicity over chirality ratio, thus $\r_+$ and $\r_-$ correspond to spectral functions of particle and plasmino modes respectively \cite{Bellac:2011kqa}. The particle mode will be used to model the left-handed half of the baryon.
As an example, we show in Fig.~\ref{fig:G0} the particle spectral function $\r_+$ as function of $\o$ for a given $p$ for timelike momenta. This is to be compared to vacuum counterpart \cite{Iqbal:2009fd}: $\r_+=2\text{sgn}(\o)\th(\o^2-p^2)\(\frac{\o+p}{\o-p}\)^{1/2}$, which shows a non-analytic singularity at $\o=p$. The presence of medium softens the singularity. The polarization of the particle mode can be obtained by tracing with the Pauli matrices
\begin{align}\label{trace_spin}
\text{tr}\big[\r_+\frac{1+\hat{p}\cdot\vec{\sigma}}{2}\s^i\big]=\r_+\hat{p}_i.
\end{align}
This is of course not surprising as it comes from spin-momentum enslavement for left-handed fermionic excitation. The right-handed counterpart is obtained by \eqref{rho_trs}. Using that the equilibrium medium is invariant under parity, we have $\r^L(\o,\vec{p})=\r^R(\o,-\vec{p})$. Noting that $\r_\pm$ are also invariant under parity, we then have
\begin{align}
\r^R=\r_+\frac{1-\hat{p}\cdot\s}{2}+\r_-\frac{1+\hat{p}\cdot\s}{2}.
\end{align}
It follows that there is no net polarization when we combine two opposite helicities to form a baryon.

Now we turn to $G_1$, which satisfies
\begin{align}\label{EOM_G1}
&G_1''+2i\frac{\omega}{f}G_1'+\frac{f'}{2f}G_1'-i\omega\frac{f'}{2f^{2}}G_1-\frac{p^{2}}{f}G_1+\frac{f'}{2f^{3/2}}\omega \frac{P\cdot\overline{\sigma}}{P^{2}}\left(G_1'+\frac{i\omega}{f}G_1\right)+\bigg[-\frac{\pd_0^X}{f}G_0'+\frac{f'}{4f^2}\pd_0^XG_0\nonumber\\
&+ip_i\frac{\pd_i^X}{f}G_0+\frac{f'}{4f^{3/2}}\frac{P\cdot\bar{\s}}{P^2}i\pd_0^X\(G_0'+\frac{2i\o}{f}G_0\)+\frac{f'}{2f^{3/2}}i\o\(\frac{D^X\cdot\bar{\s}}{-2P^2}+\frac{P\cdot D_X P\cdot\bar{\s}}{(P^2)^2}\)\(G_0'+\frac{i\o}{f}G_0\)\bigg]\nonumber\\
&=0.
\end{align}
This is an inhomogeneous ODE. The source terms involve $X$-derivatives on $G_0$. To calculate the derivatives, we first note that $G_0$ is the local bulk-boundary propagator. In the boosted AdS-Schwarzschild background, tubes of fluid velocities are related by slow-varying boost. We would obtain the same $G_0$ if we choose local rest frame in each tube. To take derivatives in a local tube, we need to express $G_0$ in neighboring tubes in the frame of local tube, to be denoted as $G_0^u$. Obviously $G_0^u$ and $G_0$ are related by an infinitesimal boost. Denoting the boost velocity by $u_i$, we have the Lorentz boost up to $O(u)$
\begin{align}\label{boost}
dt'=dt-u_idx_i,\quad dx_i'=dx_i-u_idt,
\end{align}
with primed and unprimed coordinates corresponding to local rest frames of neighboring tube and local tube. The Lorentz boost for 4-momentum is given by
\begin{align}
\o'=\o-p_iu_i,\quad p_i'-p_i-u_i\o.
\end{align}
We then have
\begin{align}
G_0^u(\o,p,z)=G_0(\o',p',z)=G_0(\o,p,z)+\frac{\pd G_0}{\pd\o}(-p_iu_i)+\frac{\pd G_0}{\pd p_i}(-u_i\o).
\end{align}
Apart from variation of fluid velocity, there is also variation of temperature. Since local temperature is fixed by $\pi T=b^{-1}=z_h^{-1}$, the variation of temperature can be achieved by scaling of coordinate $t\to\l t$, $x_i\to\l x_i$ and $z\to\l z$. The only effect of the scaling is to shift the horizon location. It follows that $G_0$ in neighboring tube with different temperature, denoted as $G_0^b$ can be expressed as infinitesimal scaling of $G_0$ in local tube
\begin{align}\label{scaling}
&G_0^b(\o,p,z)=G_0(\frac{\o}{\l},\frac{p}{\l},\l z)\nonumber\\
=&G_0(\o,p,z)+\d b\(-\o\frac{\pd}{\pd\o}-p_i\frac{\pd}{\pd p_i}+\frac{\pd}{\pd z}\)G_0(\o,p,z),
\end{align}
in the second line, we have specialized to infinitesimal variation and used $\d\l=\frac{\d b}{b}=\d b$ for our choice $b=1$ in local tube.

Now we are ready to calculate the derivative terms. Using
\begin{align}
\frac{\pd G_0}{\pd\o}=\frac{\pd A}{\pd\o}+\frac{\pd B}{\pd\o}\hat{p}\cdot\vec{\sigma},\quad \frac{\pd G_0}{\pd p_i}=\frac{\pd A}{\pd p}\hat{p}+\(\frac{\pd B}{\pd p}-\frac{B}{p}\)\hat{p}_i\hat{p}\cdot\vec{\sigma}+\frac{B}{p}\s_i,
\end{align}
we obtain the following derivative terms
\begin{align}\label{derivatives}
&\partial_{\m}^{X} G_{0}^{u}=\left(\frac{\partial A}{\partial \omega}+\frac{\partial B}{\partial \omega} \hat{p} \cdot \vec{\sigma}\right)\left(-p_{i} \partial_{\m} u_{i}\right)+\left(\frac{\partial A}{\partial p} \hat{p}_{i}+\left(\frac{\partial B}{\partial p}-\frac{B}{p}\right) \hat{p}_{i} \hat{p} \cdot \vec{\sigma}+\frac{B}{p} \sigma_{i}\right)\left(-\omega \partial_{\m} u_{i}\right)\nonumber\\
&=\(\(\frac{\pd A}{\pd\o}+\frac{\pd A}{\pd p}\frac{\o}{p}\)+\(\frac{\pd B}{\pd\o}+\(\frac{\pd B}{\pd p}-\frac{B}{p}\)\frac{\o}{p}\)\hat{p}\cdot\vec{\sigma}\)(-p_i\pd_\m u_i)+\frac{B}{p} \sigma_{i}\left(-\omega \partial_{\m} u_{i}\right).\\
&\partial_{\m}^XG_{0}^{b}=\partial_{\m}b\left(-\omega\left(\frac{\partial A}{\partial \omega}+\frac{\partial B}{\partial \omega}\hat{p}\cdot\vec{\sigma}\right)-p_{i}\left(\frac{\partial A}{\partial p}\hat{p}_{i}+\left(\frac{\partial B}{\partial p}-\frac{B}{p}\right)\hat{p}_{i}\hat{p}\cdot\vec{\sigma}+\frac{B}{p}\sigma_{i}\right)+z(A' + B'\hat{p}\cdot\vec{\sigma})\right)\nonumber\\
&=\(\(\frac{\pd A}{\pd \o}+\frac{\pd A}{\pd p}\frac{p}{\o}-\frac{zA'}{\o}\)+\(\frac{\pd B}{\pd \o}+\frac{\pd B}{\pd p}\frac{p}{\o}-\frac{zB'}{\o}\)\hat{p}\cdot\vec{\s}\)(-\o\pd_\m b).
\end{align}
We have split the terms into those with the same structure as $G_0$ and those with new structures. The possible gradients appearing in the new structure: $\pd_0u_i$, $\pd_ju_i=\e^{ijk}\o_k+\s_{ij}+\frac{1}{3}\d_{ij}\pd_k u_k$. The first one is acceleration, which induces simultaneously temperature gradient by hydrodynamic equation $\pd_ib=\pd_0u_i$ \cite{Bhattacharyya:2007vjd}. The second one includes vorticity, shear and bulk stress. We again turn off bulk stress to focus on spin polarization. The derivative terms also enter the boundary condition as
\begin{align}\label{bc_G1}
&\(\pd_z+i\o\)G_1^u=\frac{1}{2}\pd_0^XG_0^u=\big[(\cdots)1+(\cdots)\hat{p}\cdot\vec{\sigma}\big]\hat{p}_i\pd_0 u_i+\frac{B}{2p} \sigma_{i}\left(-\omega \partial_{0} u_{i}\right),\nonumber\\
&\(\pd_z+i\o\)G_1^b=\frac{1}{2}\pd_0^XG_0^b=0.
\end{align}
The second line vanishes because we have turned off the bulk stress $\pd_0b=\frac{1}{3}\pd_iu^i=0$ \cite{Bhattacharyya:2007vjd}.
From the Dirac structure of EOM \eqref{EOM_G1} and boundary condition \eqref{bc_G1}, we can deduce the Dirac structure of $G_1$ for different sources separately. For acceleration, the EOM and boundary condition are modified by new structure, which can be written schematically as
\begin{align}\label{acc_EOM}
\Box G_1=\big[(\cdots)1+(\cdots)\hat{p}\cdot\vec{\sigma}\big]\hat{p}_i\pd_0 u_i+\[C^a_1+C^a_2\hat{p}\cdot\vec{\sigma}\]\pd_0u_i\s_i,
\end{align}
with $\Box$ being short-hand notation for the operator acting on $G_1$ in \eqref{EOM_G1}. $C^a_{1,2}$ are to be evaluated from the source term in \eqref{EOM_G1} by using \eqref{derivatives}. \eqref{acc_EOM} can be solved by the following ansatz
\begin{align}\label{acc_ansatz}
G_1=\big[(\cdots)1+(\cdots)\hat{p}\cdot\vec{\sigma}\big]\hat{p}_i\pd_0 u_i+\[D^a_1+D^a_2\hat{p}\cdot\vec{\sigma}\]\pd_0u_i\s_i.
\end{align}
Note that the operator $\Box$ closes on the two structures $\big[(\cdots)1+(\cdots)\hat{p}\cdot\vec{\sigma}\big]$ and $\[D_1^a+D_2^a\hat{p}\cdot\vec{\sigma}\]\pd_0u_i\s_i$ separately. It follows that we can solve for the two parts separately.
The boundary condition for this case is determined by merging the two derivatives from acceleration and temperature gradient in \eqref{derivatives} as
\begin{align}
(\pd_z+i\o)G_1=\[(\cdots)1+(\cdots)\hat{p}\cdot\vec{\sigma}\]\hat{p}_i\pd_0 u_i-\frac{\bar{B}}{2}\pd_0u_i\s_i,
\end{align}
with $\bar{B}=\frac{\o}{p}B$.

We will not solve for the structure $\[(\cdots)1+(\cdots)\hat{p}\cdot\vec{\sigma}\]\hat{p}_i\pd_0 u_i$ in $G_1$ for the following reason: the spin dependent term $(\cdots)\hat{p}\cdot\vec{\s}\hat{p}_i\pd_0u_i$ is parity odd. By using \eqref{rho_trs}, we deduce an opposite contribution from the right-handed counterpart. It follows that the correction cancels out in spin polarization of baryon. In contrast, the spin independent term $\hat{p}_i\pd_0u_i$ does add up. We will focus on the spin dependent contributions. This allows us to drop the structure $\[(\cdots)1+(\cdots)\hat{p}\cdot\vec{\sigma}\]\hat{p}_i\pd_0 u_i$ and work with a simplified EOM
\begin{align}\label{acc_EOM_simp}
\Box G_1=\[C^a_1+C^a_2\hat{p}\cdot\vec{\sigma}\]\pd_0u_i\s_i,
\end{align}
with
\begin{align}\label{acc_bc}
(\pd_z+i\o)G_1=-\frac{\bar{B}}{2}\pd_0u_i\s_i.
\end{align}
$G_1$ can be solved by
\begin{align}\label{acc_ansatz_simp}
G_1=\[D^a_1+D^a_2\hat{p}\cdot\vec{\sigma}\]\pd_0u_i\s_i.
\end{align}

To proceed, we split the differential operator as $\Box=\Box_1+\Box_2\hat{p}\cdot\vec{\s}$. Plugging \eqref{acc_ansatz_simp} into \eqref{acc_EOM_simp}, we obtain
\begin{align}\label{EOM_D}
&\Box_1 D_1^a+\Box_2 D_2^a=C_1^a,\nonumber\\
&\Box_1 D_2^a+\Box_2 D_1^a=C_2^a.
\end{align}
The boundary condition from \eqref{acc_bc} reduces to
\begin{align}\label{bc_D}
(\pd_z+i\o)D_1^a=-\frac{\bar{B}}{2},\quad (\pd_z+i\o)D_2^a=0.
\end{align}
The other boundary condition at the horizon is fixed as follows: by analyzing the asymptotic behavior of $G_1$, we find two independent series in the form of $(1-z)^{-i\o}$ and powers of $(1-z)$ (including also logarithms). The former corresponds to outfalling solution and the latter is a variation of the regular solution. We keep the latter. This fixes the solution up to two independent integration constants to be fixed by the boundary condition \eqref{bc_D} at $z=0$. In fact we do not have to tune the integration constants: a new inhomogeneous solution can be obtained by adding homogeneous solutions, which can be chosen as $A\pd_0u_i\s_i+B\hat{p}\cdot\vec{\s}\pd_0u_i\s_i$. Crucially, unlike in the solution of $G_0$, here $A$ and $B$ can satisfy arbitrary boundary condition at $z=0$, so the two independent homogeneous solutions precisely match the number of integration constants. In practice, we integrate \eqref{EOM_D} using arbitrary integration constants from the horizon, then find suitable homogeneous solution to be added such that \eqref{bc_D} is satisfied.

The cases of shear is analyzed similarly. The counterpart of \eqref{acc_EOM} is given by
\begin{align}\label{shear_EOM}
\Box G_1=\big[(\cdots)1+(\cdots)\hat{p}\cdot\vec{\sigma}\big]\hat{p}_i\hat{p}_j\s_{ij}+\[C^a_1+C^a_2\hat{p}\cdot\vec{\sigma}\]\hat{p}_j\s_{ij}\s_i.
\end{align}
Since shear arises from $\pd_i^XG_0^u$, the boundary condition for the shear case follows from \eqref{bc_G1} trivially as
\begin{align}\label{shear_bc}
(\pd_z+i\o)G_1=0.
\end{align}
We can use the same argument as in the acceleration case. The spin dependent structure $\hat{p}\cdot\vec{\sigma}\hat{p}_i\hat{p}_j\s_{ij}$ is parity odd, so its contribute to spin polarization will be canceled by an opposite contribution from right-handed counterpart due to \eqref{rho_trs}. Thus we can drop the structure $\big[(\cdots)1+(\cdots)\hat{p}\cdot\vec{\sigma}\big]\hat{p}_i\hat{p}_j\s_{ij}$ and use a simplfied ansatz for $G_1$ as
\begin{align}
G_1=\big[D_1^\s+D_2^\s\hat{p}\cdot\vec{\s}\big]\hat{p}_j\s_{ij}\s_i.
\end{align}

One might hope similar simplification can be used for the case of vorticity. It turns out not to be the case. The source terms in \eqref{EOM_G1} can be written as
\begin{align}\label{vorticity_source}
&E_1\e^{ijk}\o_i\hat{p}_j\s_k+E_2\hat{p}\cdot\vec{\sigma}\e^{ijk}\o_i\hat{p}_j\s_k+E_3\s_j\e^{ijk}\o_k\s_i+E_4\hat{p}\cdot\vec{\sigma}\s_j\e^{ijk}\o_k\s_i\nonumber\\
=&i\[(E_1-2E_4)\hat{p}\cdot\vec{\s}\o\cdot\vec{\s}-E_1\hat{p}\cdot\o+(E_2-2E_3)\o\cdot\vec{\s}-E_2\hat{p}\cdot\vec{\s}\hat{p}\cdot\o\]\nonumber\\
&\equiv\[C_1^\o+C_2^\o\hat{p}\cdot\vec{\s}\]\o_i\s_i+\[C_1^{\o_\pr}+C_2^{\o_\pr}\hat{p}\cdot\vec{\s}\]\hat{p}\cdot\vec{\o}.
\end{align}
The resulting EOM is given by
\begin{align}
\Box G_1=\[C_1^\o+C_2^\o\hat{p}\cdot\vec{\s}\]\o_i\s_i+\[C_1^{\o_\pr}+C_2^{\o_\pr}\hat{p}\cdot\vec{\s}\]\hat{p}\cdot\vec{\o},
\end{align}
with the same boundary condition as \eqref{shear_bc}. This is to be solved by
\begin{align}\label{vorticity_ansatz}
G_1=\[D_1^\o+D_2^\o\hat{p}\cdot\vec{\s}\]\o_i\s_i+\[D_1^{\o_\pr}+D_2^{\o_\pr}\hat{p}\cdot\vec{\s}\]\hat{p}\cdot\vec{\o}.
\end{align}
Crucially, $\hat{p}\cdot\vec{\s}\hat{p}\cdot\vec{\o}$ is parity even, unlike the other two cases. It follows that both structures contribute to the spin polarization of the baryon upon combining the left and right-handed components.

\begin{figure}
	\centering
	\includegraphics[width=0.48\linewidth]{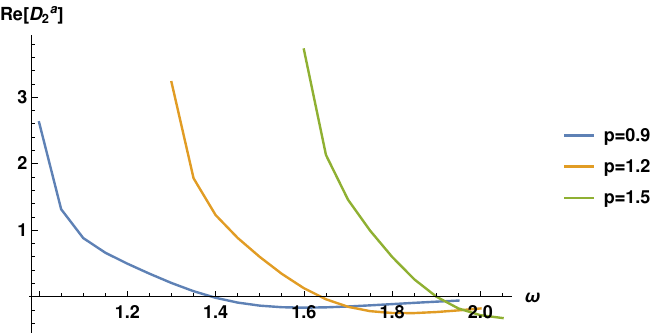}
	\includegraphics[width=0.48\linewidth]{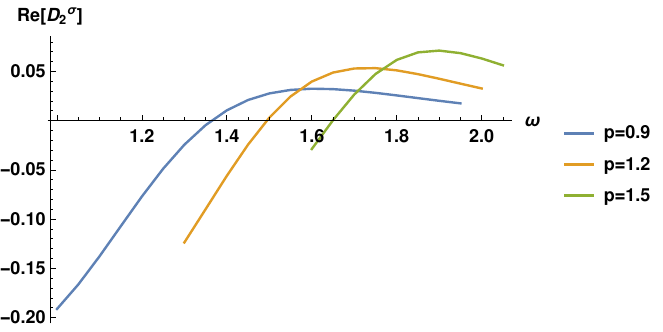}
	\includegraphics[width=0.48\linewidth]{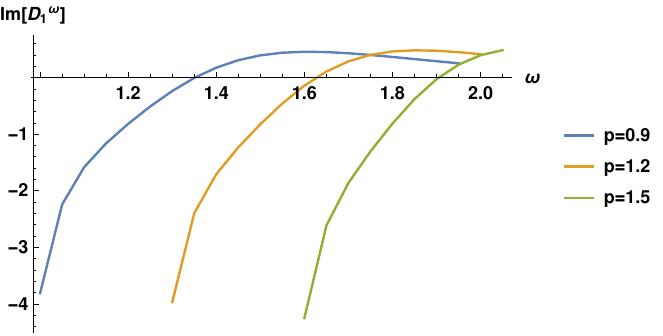}
	\includegraphics[width=0.48\linewidth]{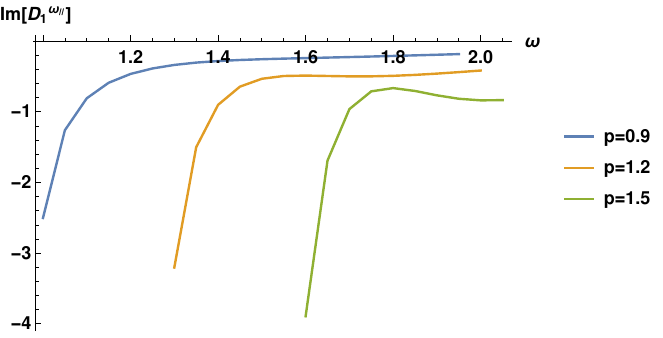}
	\caption{The coefficients of gradient correction defined in \eqref{probe_combined} for local equilibrium state versus $\o$ for three values of $p$. The unit is set by $\p T=1$. The magnitude of coefficients generically grow as the lightcone is approached $\o\to p$, which is reminiscent of the singular behavior in vacuum correlator. The coefficients are suppressed at large $\o$. }
	\label{fig:G1_probe}
\end{figure}

With $G_1$ solved, the resulting boundary correlator is given by
\begin{align}\label{probe_G}
\d G_R&=\[D_1^a+D_2^a\hat{p}\cdot\vec{\sigma}\]\pd_0u_i\s_i+\[D_1^\s+D_2^\s\hat{p}\cdot\vec{\sigma}\]\hat{p}_j\s_{ij}\s_i+\[D_1^\o+D_2^\o\hat{p}\cdot\vec{\sigma}\]\o_i\s_i\nonumber\\
&+\[D_1^{\o_\pr}+D_2^{\o_\pr}\hat{p}\cdot\vec{\sigma}\]\hat{p}\cdot\vec{\o},
\end{align}
where it is understood that the limit $z\to0$ is taken.
It is instructive to extract the polarization dependence by tracing \eqref{probe_G} with $\s_k$ to give polarization of the spectral function from \eqref{rho_GR}
\begin{align}\label{rho_probe}
	&\frac{1}{2}\text{tr}\[\d \r^L\s_k\]=
	2\bigg[\(\text{Im}[D_1^a]\pd_0u_k-\text{Re}[D_2^a]\e^{ijk}\hat{p}_j\pd_0u_i\)+\(\text{Im}[D_1^\s]\hat{p}_j\s_{kj}-\text{Re}[D_2^\s]\e^{ijk}\hat{p}_j\hat{p}_l\s_{il}\)\nonumber\\
	&+\(\text{Im}[D_1^\o]\o_k-\text{Re}[D_2^\o]\e^{ijk}\hat{p}_j\o_i\)+\text{Im}[D_1^{\o_\pr}]\o_k^\pr\bigg],
\end{align}
with $\o_k^\pr\equiv\hat{p}_k\hat{p}\cdot\o$.
We have reinstated the superscript $L$ for left-handed fermionic operator. The counterpart for right-handed operator is obtained by parity transformation \eqref{rho_trs}. Note that $\d\r^{L/R}$ arise from $D_{\text{probe}}^{(1)}$, which ultimately reduces to the structures $\pd_0u_k$, $\e^{ijk}\hat{p}_j\pd_0u_i$ etc. Applying parity transformation to these structures, we obtain
\begin{align}
	&\frac{1}{2}\text{tr}\[\d \r^R\s_k\]=
	2\bigg[\(-\text{Im}[D_1^a]\pd_0u_k-\text{Re}[D_2^a]\e^{ijk}\hat{p}_j\pd_0u_i\)+\(-\text{Im}[D_1^\s]\hat{p}_j\s_{kj}-\text{Re}[D_2^\s]\e^{ijk}\hat{p}_j\hat{p}_l\s_{il}\)\nonumber\\
	&+\(\text{Im}[D_1^\o]\o_k+\text{Re}[D_2^\o]\e^{ijk}\hat{p}_j\o_i\)+\text{Im}[D_1^{\o_\pr}]\o_k^\pr\bigg].
\end{align}
Combining contributions from both helicities, we obtain
\begin{align}\label{probe_combined}
	&\frac{1}{2}\text{tr}\[(\d \r^L+\d\r^R)\s_k\]=4\(-\text{Re}[D_2^a]\e^{ijk}\hat{p}_j\pd_0u_i-\text{Re}[D_2^\s]\e^{ijk}\hat{p}_j\hat{p}_l\s_{il}+\text{Im}[D_1^{\o}]\o_k+\text{Im}[D_1^{\o_\pr}]\o_k^\pr\).
\end{align}
We show in Fig.~\ref{fig:G1_probe} numerical results for the coefficients in \eqref{probe_combined} for time-like momenta for three values of $p$.

\subsection{Gradient correction to the Dirac operator}

Now we consider gradient correction to the Dirac operator, which encodes gradient correction to the medium density matrix $D_\text{fluid}^{(1)}$
\begin{align}
\G^M\nabla_M\ps=\G^M\(\pd_M+\frac{1}{4}\o_{Mab}\g^{ab}\)\ps=0,
\end{align}
with
\begin{align}\label{spin_connection}
\omega_{Mab}=\eta_{cb}\left(e_{a}^{N}\partial_{M}e_{N}^{c}-e_{a}^{N}\Gamma_{MN}^{K}e_{K}^{c}\right).
\end{align}
There are two possible corrections to the Dirac operator at $O(\pd_X)$: one from obvious $O(\pd_X)$ metric correction, which can be attributed to steady state effect of the medium, and the other less obvious one from the derivative term $\h_{cb}e_a^N\pd_M e_N^c$ on $O(\pd_X^0)$ vielbein in $\o_{Mab}$. The form of the correction in the latter case suggests it originates from choice of local Lorentz frame, which exists in local equilibrium state of the medium. Following the logic, we shall denote the corresponding medium density matrices as $D_\text{fluid}^{(1)\text{ss}}$ and $D_\text{fluid}^{(1)\text{le}}$ respectively.
We start with calculation of the latter contribution. We define $\d\o_{ab}=\h_{cb}e_a^N\pd_M \d e_N^c$ with $\d e_N^c$ coming from variation of either fluid velocity or temperature. This way we have $\h_{cb}e_a^N\pd_M e_N^c=\frac{\d\o_{ab}}{\d x^M}$ with $M$ taking only boundary coordinates. 
The vielbeins are chosen as\footnote{We stress that the following choice is not in contradictory with the asymptotic condition of vielbein \eqref{vielbein_EF} written in a fluid tube. The condition is satisfied in local rest frame of a given tube.}
\begin{align}\label{vielbeins}
&e^t_{\hat{t}}=\frac{z}{f^{1/2}},\;e^t_{\hat{z}}=-\frac{z}{f^{1/2}},\;e^i_{\hat{i}}=z,\;e^z_{\hat{z}}=z f^{1/2},\; e^t_{\hat{i}}=u_i z,\; e^i_{\hat{t}}=- e^i_{\hat{z}}=u_i\frac{z}{f^{1/2}},\nonumber\\
&e^{\hat{t}}_t=\frac{f^{1/2}}{z},\;e^{\hat{t}}_z=\frac{1}{zf^{1/2}},\;e^{\hat{i}}_i=\frac{1}{z},\;e^{\hat{z}}_z=\frac{1}{zf^{1/2}},\;e^{\hat{t}}_i=-u_i\frac{f^{1/2}}{z},\;e^{\hat{i}}_t=-\frac{u_i}{z},
\end{align}
where we have only kept up to $O(u_i)$ for our purpose. We then have the following $\d\o_{ab}$
\begin{align}\label{domega}
\d\o_{\hat{i}\hat{t}}=u_if^{1/2},\;\d\o_{\hat{t}\hat{i}}=-\frac{u_i}{f^{1/2}},\;\d\o_{\hat{z}\hat{i}}=\frac{u_i}{f^{1/2}},\;\d\o_{\hat{z}\hat{t}}=\frac{\d f}{f}.
\end{align}
The gradient correction to the Dirac operator $\d(\G^M\nabla_{M})$ is then given by
\begin{align}\label{dDirac1}
&\frac{1}{4}\left[z\partial_{0}u_{i}\sigma_{i}
\begin{pmatrix}-1&-i\\i&-1
\end{pmatrix}
+z\sigma_{j}\sigma_{i}\partial_{j}u_{i}
\begin{pmatrix}-1&i(1 + f)\\i(1 + f)&1
\end{pmatrix}
\frac{1}{f^{1/2}}-z^{2}\frac{f'}{f}\frac{\partial_{i}b}{b}
\begin{pmatrix}
-\sigma_{i}&0\\0&-\sigma_{i}
\end{pmatrix}\right]\nonumber\\
=&\frac{1}{4}\left[z\partial_{0}u_{i}\sigma_{i}
\begin{pmatrix}-1&-i\\i&-1
\end{pmatrix}
+z2i\omega_{i}\sigma_{i}
\begin{pmatrix}
-1&i(1 + f)\\i(1 + f)&1
\end{pmatrix}
\frac{1}{f^{1/2}}-z^{2}\frac{f'}{f}\partial_{0}u_{i}
\begin{pmatrix}-\sigma_{i}&0\\0&-\sigma_{i}
\end{pmatrix}\right],
\end{align}
where we have used $\frac{\pd_ib}{b}=\pd_0u_i$ and $\pd_ju_i\s_j\s_i=-2i\o_k\s_k$ from the vanishing of bulk stress. Note that the contribution from shear stress drops out at this step. We have also dropped temporal derivative of temperature in \eqref{dDirac1} as it is related to bulk stress as $\pd_0b=\pd_ku_k$ \cite{Bhattacharyya:2007vjd}. Note that the correction can be treated as a constant in a given fluid element, so that the Wigner transform essentially reduces to the Fourier transform. We then easily obtain the following EOM
\begin{align}\label{EOM_RL2}
&-P\cdot\sigma\chi_{L}+\frac{i\omega}{f^{1/2}}\chi_{R}+f^{1/2}\chi_{R}'-\frac{i}{4}\partial_{0}u_{i}\sigma_{i}z\chi_{L}-\frac{1}{4}\partial_{0}u_{i}\sigma_{i}(z - z^{2}\frac{f'}{f})\chi_{R}-\frac{z}{2f^{1/2}}\omega_{i}\sigma_{i}(\chi_{L}+i\chi_{R})\nonumber\\
&-\frac{z f^{1/2}}{2}\omega_{i}\sigma_{i}\chi_{L}=0,\\
&-P\cdot\overline{\sigma}\chi_{R}-\frac{i\omega}{f^{1/2}}\chi_{L}-f^{1/2}\chi_{L}'+\frac{1}{4}\partial_{0}u_{i}\sigma_{i}z\chi_{R}-\frac{1}{4}\partial_{0}u_{i}\sigma_{i}(z - z^{2}\frac{f'}{f})\chi_{L}-\frac{z}{2f^{1/2}}\omega_{i}\sigma_{i}(-i\chi_{L}+\chi_{R})\nonumber\\
&-\frac{z f^{1/2}}{2}\omega_{i}\sigma_{i}\chi_{R}=0.
\end{align}
As before, we shall derive one second order ODE from the two first order ODE. We do so separately for the sources of acceleration and vorticity. For acceleration, we find the following EOM for $G$ (from the counterpart for $\c_R$) as
\begin{align}
&\Box G_{1}+\bigg[\frac{i p_{i}\partial_{0}u_{i}zG_{0}}{2f}-\frac{\partial_{0}u_{i}\sigma_{i}G_{0}}{4\sqrt{f}}+\frac{\partial_{0}u_{i}\sigma_{i}zG_{0}f'}{2f^{3/2}}-\frac{\partial_{0}u_{i}\sigma_{i}\sigma_{i}zG_{0}f'}{2f^{3/2}}+\frac{i\partial_{0}u_{i}\sigma_{i}P\cdot\overline{\sigma}(i\omega G_{0}+fG_{0}')}{4fP^{2}}\nonumber\\
&-\omega\frac{f'}{2f^{2}}\left(\frac{-P\cdot\overline{\sigma}\left(-\frac{1}{4}\partial_{0}u_{i}\sigma_{i}\left(z - z^{2}\frac{f'}{f}\right)G_{0}\right)}{P^{2}}+\frac{\left(i\frac{\partial_{0}u_{i}\sigma_{i}z}{4P^{2}}-\frac{i p_{i}\partial_{0}u_{i}zP\cdot\overline{\sigma}}{2P^{4}}\right)(i\omega G_{0}+fG_{0}')}{\sqrt{f}}\right)\bigg]=0
\end{align}
Following the procedure as the previous sections, we do not keep terms with the same Dirac structure as $G_0$ in the source. These include terms proportional to $p_i\pd_0u_i$. Using
\begin{align}
\pd_0u_i\s_i\hat{p}\cdot\vec{\s}=2\hat{p}_i\pd_0u_i-\hat{p}\cdot\vec{\s}\pd_0u_i\s_i,
\end{align}
and again dropping terms proportional to $\hat{p}_i\pd_0u_i$, the EOM can be written as
\begin{align}\label{acc_EOM2}
\Box G_1=\[C^a_3+C^a_4\hat{p}\cdot\vec{\sigma}\]\pd_0u_i\s_i.
\end{align}
It is formally the same as \eqref{acc_EOM_simp}, but with a slightly different boundary condition, which we now derive. Applying Fourier transform to \eqref{bb_prop}, and using \eqref{EOM_RL2}, we obtain
\begin{align}\label{bc2_acc_derivation}
\(P\cdot\s+\frac{i}{4}\pd_0u_i\s_iz\)\c_L(\o,\vec{p},z)=\(f^{1/2}\pd_z+\frac{i\o}{f^{1/2}}-\frac{1}{4}\pd_0u_i\s_i\(z-z^2\frac{f'}{f}\)\)G(\o,\vec{p},z)\c_L^b(\o,\vec{p}).
\end{align}
Taking $z\to0$, we simply find $P\cdot\s=\(\pd_z+i\o\)G(\o,\vec{p},z\to0)$. The same boundary condition has been satisfied by $G_0$, so that we have
\begin{align}\label{bc2_acc}
\(\pd_z+i\o\)G_1(\o,\vec{p},z\to0)=0.
\end{align}

We will not repeat the derivation for the case of vorticity. The resulting EOM is given by
\begin{align}
&\Box G_{1}+\bigg[\frac{z\omega\omega_i\sigma_i G_{0}}{f^{2}}+\frac{i\e^{ijk}\omega_i p_j\sigma_k G_{0}}{f^{3/2}}(1 + f)-\frac{i\omega_i\sigma_i G_{0}}{2f}+\frac{i z\omega_i\sigma_i G_{0}f'}{4f^{2}}-\frac{i z\omega_i\sigma_i G_{0}'}{f}+\nonumber\\
&\frac{\omega_i\sigma_i P\cdot\overline{\sigma}(i\omega G_{0}+fG_{0}')}{2f^{3/2}P^{2}}
+\frac{\omega_i\sigma_i P\cdot\overline{\sigma}(i\omega G_{0}+fG_{0}')}{2\sqrt{f}P^{2}}-\frac{z\omega_i\sigma_i P\cdot\overline{\sigma}(i\omega G_{0}+fG_{0}')f'}{4f^{5/2}P^{2}}(1 - f)\nonumber\\
&-\omega\frac{f'}{2f^{2}}\left(\frac{i zP\cdot\overline{\sigma}\omega_i\sigma_i G_{0}}{2\sqrt{f}P^{2}}+\frac{1}{\sqrt{f}}\left(-\frac{z\omega_i\sigma_i(1 + f)}{2\sqrt{f}P^{2}}+\frac{i z\e^{ijk}\o_ip_j\s_k (1 + f)P\cdot\overline{\sigma}}{\sqrt{f}P^{4}}\right)\right)\bigg]=0,
\end{align}
Again we need to treat the vorticity case carefully, which is expected to lead to two independent contributions to polarization. Using $i\e^{ijk}\o_ip_j\s_k=\vec{p}\cdot\vec{\o}-\vec{p}\cdot\vec{\s}\o_i\s_i=-\vec{p}\cdot\vec{\o}+\o_i\s_i\vec{p}\cdot\vec{\s}$,
we can rewrite the EOM as
\begin{align}
\Box G_1=\[C^\o_3+C^\o_4\hat{p}\cdot\vec{\sigma}\]\o_i{\sigma}_i+\[C^{\o_\pr}_3+C^{\o_\pr}_4\hat{p}\cdot\vec{\sigma}\]\hat{p}_i\o_i.
\end{align}
The boundary condition is derived similarly as \eqref{bc2_acc_derivation} to give
\begin{align}\label{bc2_omega}
\(\pd_z+i\o\)G_1(\o,\vec{p},z\to0)=0.
\end{align}

Finally we consider the steady state correction from metric at $O(\pd_X)$, which can modify the spin connection through either the vielbein or the affine connection. The following components of vielbein will be corrected
\begin{align}
\d e^{\hat{i}}_j=\frac{F}{z}\s_{ij},\; \d e^{\hat{t}}_i=\d e^{\hat{z}}_i=-\frac{1}{f^{1/2}}\pd_0u_i,
\end{align}
which give rise to the following coupled EOM
\begin{align}\label{EOM_RL3}
&-\left(P\cdot\sigma+\frac{1}{2}p_j\s_{ij}\s_i F+\left(\frac{i}{2}-i\frac{z f'}{4f}\right)\partial_{0}u_{i}\sigma_{i}\right)\chi_{L}+\left(\frac{1}{4}-\frac{z f'}{4f}\right)\partial_{0}u_{i}\sigma_{i}\chi_{R}-iz\partial_{0}u_{i}\sigma_{i}\chi_{L}' \nonumber\\
& +i\frac{\omega}{f^{1/2}}\chi_{R}+f^{1/2}\chi_{R}' = 0
,\nonumber\\
&-\left(P\cdot\overline{\sigma}-\frac{1}{2}p_j\s_{ij}\s_i F-\left(\frac{i}{2}-i\frac{z f'}{4f}\right)\partial_{0}u_{i}\sigma_{i}\right)\chi_{R}+\left(\frac{1}{4}-\frac{z f'}{4f}\right)\partial_{0}u_{i}\sigma_{i}\chi_{L}+iz\partial_{0}u_{i}\sigma_{i}\chi_{R}' \nonumber\\
&-i\frac{\omega}{f^{1/2}}\chi_{L}-f^{1/2}\chi_{L}' = 0.
\end{align}
We will decouple the equations for the source of shear and acceleration separately. For the shear, we find
\begin{align}\label{shear_EOM3}
\Box G_1+\frac{\sigma_{ij}\sigma_{i}}{4f^{5/2}P^{2}}p_{j}\left(-i\omega^{2}FG_{0}f'+2i\omega f^{3/2}P\cdot\overline{\sigma}G_{0}F'-\omega fFf'G_{0}'+2f^{5/2}P\cdot\overline{\sigma}F'G_{0}'\right)=0,
\end{align}
with the following boundary condition
\begin{align}\label{bc3_shear}
\(\pd_z+i\o\)G_1(\o,\vec{p},z\to0)=0.
\end{align}
For the acceleration, the decoupling is complicated by the presence of terms $iz\pd_0u_i\s_i\c_{L/R}'$ in \eqref{EOM_RL3}. Nevertheless the decoupling can be done perturbatively in the source. The resulting EOM is lengthy and not shown explicitly. The corresponding boundary condition is derived as
\begin{align}\label{bc3_acc}
&\(p\cdot\s+\frac{i}{2}\pd_0u_i\s_i\)=\(\pd_z+i\o+\frac{1}{4}\pd_0u_i\s_i\)G(\o,\vec{p},z\to0),\nonumber\\
&\Rightarrow\;\frac{i}{2}\pd_0u_i\s_i=\(\pd_z+i\o\)G_1(\o,\vec{p},z\to0)+\frac{1}{4}\pd_0u_i\s_iG_0(\o,\vec{p},z\to0).
\end{align}

\subsection{Retarded correlator from gradient correction to the medium}

\begin{figure}
	\centering
	\includegraphics[width=0.48\linewidth]{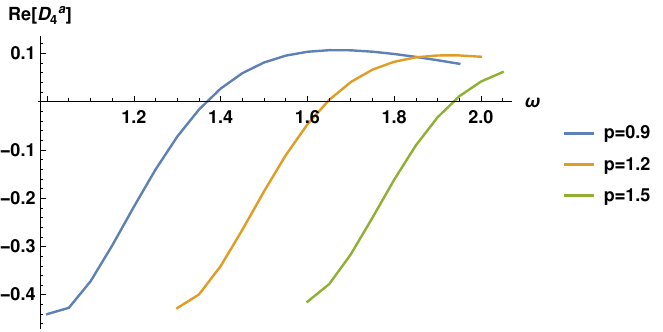}
	\includegraphics[width=0.48\linewidth]{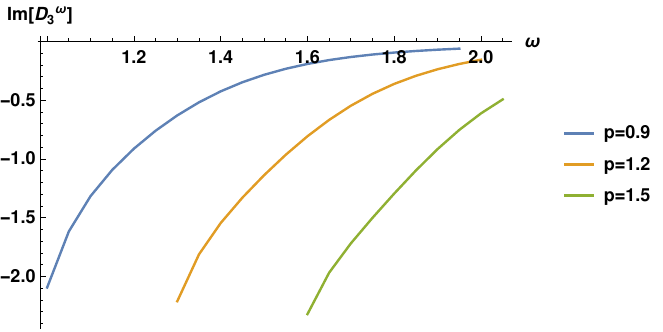}
	\includegraphics[width=0.48\linewidth]{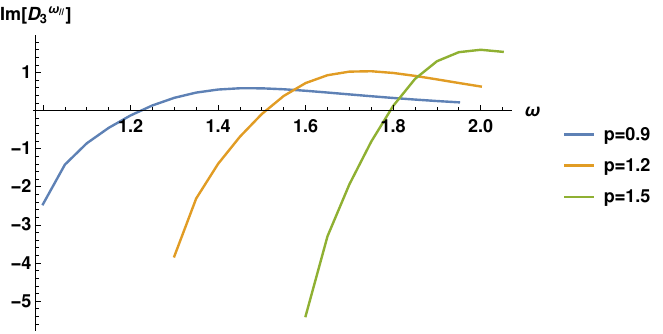}
	\includegraphics[width=0.48\linewidth]{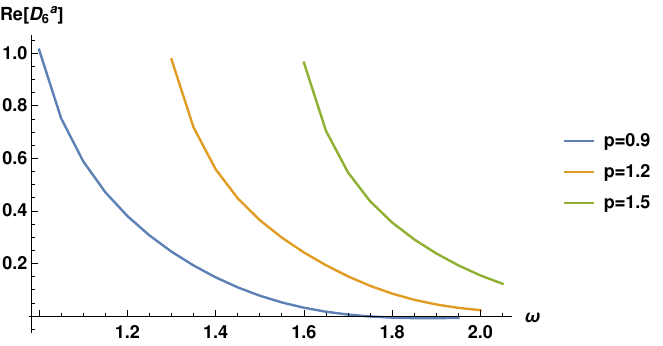}
	\includegraphics[width=0.48\linewidth]{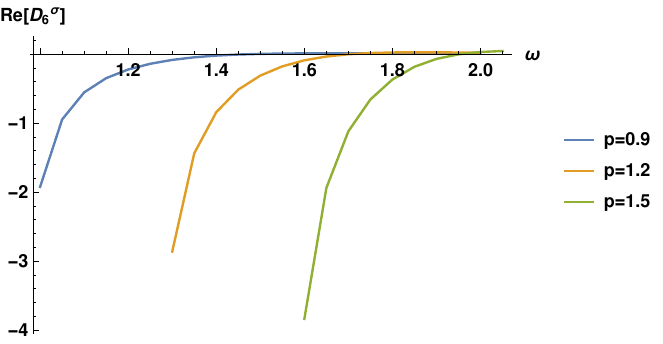}
	\caption{The coefficients of gradient correction defined in \eqref{medium_G} versus $\o$ for three values of $p$. The unit is set by $\p T=1$. The first three and last two plots arise from $D_{\text{fluid}}^{(1)\text{le}}$ and $D_{\text{fluid}}^{(1)\text{ss}}$ respectively, corresponding to gradient correction to local equilibrium state and steady state effect of the fluid. The spectral function is enhanced toward the lightcone and suppressed at large $\o$.}
	\label{fig:G1_medium}
\end{figure}

Now we solve the corresponding EOMs and extract the boundary correlator. The procedure is similar to subsection.~\ref{sec:probe}. The following ansatz will be used
\begin{align}\label{medium_G}
&G_1=(D_3^a+\hat{p}\cdot\vec{\s}D_4^a)\pd_0u_i\s_i,\nonumber\\
&G_1=(D_3^\o+\hat{p}\cdot\vec{\s}D_4^\o)\o_i\s_i+(D_3^{\o_\pr}+\hat{p}\cdot\vec{\s}D_4^{\o_\pr})\hat{p}_i\o_i,\nonumber\\
&G_1=(D_5^a+\hat{p}\cdot\vec{\s}D_6^a)\pd_0u_i\s_i,\nonumber\\
&G_1=(D_5^\s+\hat{p}\cdot\vec{\s}D_6^\s)\hat{p}_j\s_{ij}\s_i,
\end{align}
with the subscripts $3,4$ and $5,6$ corresponding to local equilibrium and steady state effect respectively as argued below \eqref{spin_connection}. Collecting all gradient corrections to the Dirac operator, we can extract the polarization dependence of spectral function as
\begin{align}\label{rho_medium}
&\frac{1}{2}\text{tr}[\d\r^L\s_k]=2\bigg[\text{Im}[D_3^a+D_5^a]\pd_0u_k-\text{Re}[D_4^a+D_6^a]\e^{ijk}\hat{p}_j\pd_0u_i\nonumber\\
&+\text{Im}[D_5^\s]\hat{p}_j\s_{kj}-\text{Re}[D_6^\s]\e^{ijk}\hat{p}_j\hat{p}_l\s_{il}+\text{Im}[D_3^\o]\o_k-\text{Re}[D_4^\o]\e^{ijk}\hat{p}_j\o_i+\text{Im}[D_3^{\o_\pr}]\o_k^\pr\bigg].
\end{align}
We have reinstated the superscript $L$ for left-handed operator. The counterpart for right-handed operator is obtained by \eqref{rho_trs} as before
\begin{align}
&\frac{1}{2}\text{tr}[\d\r^R\s_k]=2\bigg[-\text{Im}[D_3^a+D_5^a]\pd_0u_k-\text{Re}[D_4^a+D_6^a]\e^{ijk}\hat{p}_j\pd_0u_i\nonumber\\
&-\text{Im}[D_5^\s]\hat{p}_j\s_{kj}-\text{Re}[D_6^\s]\e^{ijk}\hat{p}_j\hat{p}_l\s_{il}+\text{Im}[D_3^\o]\o_k+\text{Re}[D_4^\o]\e^{ijk}\hat{p}_j\o_i+\text{Im}[D_3^{\o_\pr}]\o_k^\pr\bigg].
\end{align}
The combined spectral function for baryon is given by
\begin{align}
\frac{1}{2}\text{tr}[(\d\r^L+\d\r^R)\s_k]=4\bigg[-\text{Re}[D_4^a+D_6^a]\e^{ijk}\hat{p}_j\pd_0u_i-\text{Re}[D_6^\s]\e^{ijk}\hat{p}_j\hat{p}_l\s_{il}+\text{Im}[D_3^\o]\o_k+\text{Im}[D_3^{\o_\pr}]\o_k^\pr\bigg].
\end{align}
We show in Fig.~\ref{fig:G1_medium} the coefficient functions versus $\o$ for three values of $p$.

\subsection{Summary of gradient corrections and discussion}

\begin{figure}
	\centering
	\includegraphics[width=0.48\linewidth]{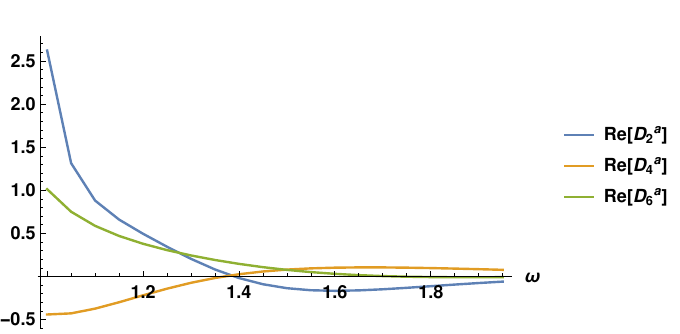}
	\includegraphics[width=0.48\linewidth]{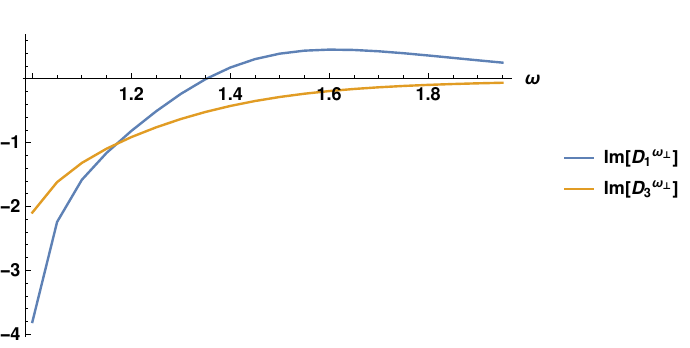}
	\includegraphics[width=0.48\linewidth]{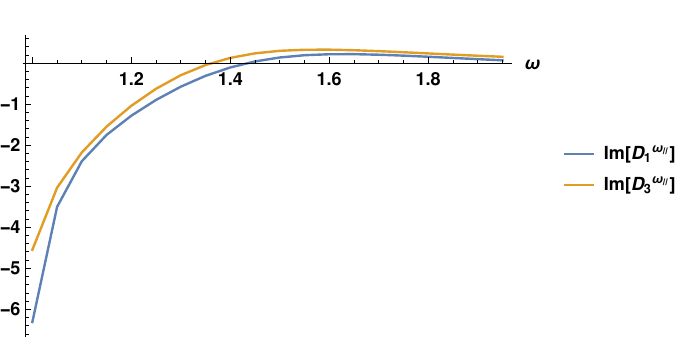}	\includegraphics[width=0.48\linewidth]{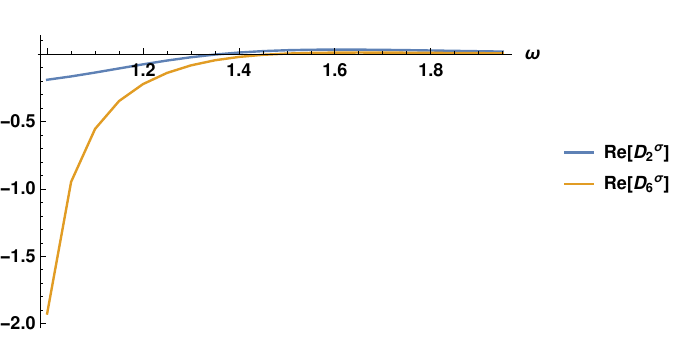}
	\caption{Comparisons of contributions from the same hydrodynamic gradients listed in Table.~\ref{tab:summary} as a function of $\o$ at $p=0.9$. The unit is set by $\p T=1$. We have defined $\o_k^\pp=\o_k-\o_k^\pr$ and switched to the basis $\o_\pp$ and $\o_\pr$.}
	\label{fig:summary}
\end{figure}

We summarize gradient contributions to the polarization dependence of the spectral function in Table.~\ref{tab:summary}.
\begin{table}[h]\label{tab:summary}
	\centering
	\begin{tabular}{c|c|c|c|c}
		\hline
		$\text{tr}[(\d\r^L+\d\r^R)\s_k]$& $\e^{ijk}\hat{p}_j\pd_0u_i$&  $\e^{ijk}\hat{p}_j\hat{p}_l\s_{il}$& $\o_k^\pp$& $\o_k^\pr$\\
		\hline
		$D_\text{probe}^{(1)}$& $-4\text{Re}[D_2^a]$& $-4\text{Re}[D_2^\s]$& $4\text{Im}[D_1^\o]$& $4\text{Im}[D_1^\o+D_1^{\o_\pr}]$\\
		$D_\text{fluid}^{(1)\text{le}}$& $-4\text{Re}[D_4^a]$& & $4\text{Im}[D_3^\o]$& $4\text{Im}[D_3^\o+D_3^{\o_\pr}]$\\
		$D_\text{fluid}^{(1)\text{ss}}$& $-4\text{Re}[D_6^a]$&  $-4\text{Re}[D_6^\s]$& &\\
		\hline
	\end{tabular}
	\caption{Summary of gradient correction. The leftmost column denotes the origin of the polarized spectral function. $D_\text{probe}^{(1)}$ comes from gradient correction to density matrix of the probe baryon. $D_\text{fluid}^{(1)\text{le}}$ and $D_\text{fluid}^{(1)\text{ss}}$ come from gradient correction to density matrices in local equilibrium state and  steady state of the fluid respectively. We have defined $\o_k^\pp=\o_k-\o_k^\pr$ and switched to the basis $\o_\pp$ and $\o_\pr$.}
\end{table}
We show in Fig.~\ref{fig:summary} comparisons of contributions from the same hydrodynamic gradients listed in Table.~\ref{tab:summary} as a function of $\o$ for a fixed $p$.

It is instructive to compare the present results with counterpart obtained in the weak coupling regime. In \cite{Lin:2024svh}, polarized spectral function for quarks from gradient corrections has been obtained based on quantum kinetic approach appropriate for weak coupling, with the polarized part of the spectral function scaling as $O(g^2C_F)\sim O(\l)$ as $N_c\to\infty$.
In the strong coupling regime, we find the polarized part of the spectral function scales as $O(\l^0)$ from the holographic model. We stress that the scaling comparison above is meaning as the gradient corrections to spectral function have been normalized by equilbrium counterpart, so are independent of being from quarks or baryons. 

With possible gradient corrections classified in terms of density matrices, we may try to make a closer comparison. We first note that the polarized part of the spectral function found in weak coupling regime include $D_\text{probe}^{(1)}$ and $D_\text{fluid}^{(1)\text{le}}$, but not $D_\text{fluid}^{(1)\text{ss}}$\footnote{$D_\text{fluid}^{(1)\text{ss}}$ in kinetic description involves shifted distribution function in steady state that is considered in \cite{Lin:2022tma,Lin:2024zik}, but the corresponding correction to spectral function has not be worked out.}. Recall the weakly coupled results arise from \cite{Lin:2024svh}: (i) $O(\pd_X^0)$ local equilibrium self-energy combined with $O(\pd_X)$ correction to the resummation equation (Kadanoff-Baym equation); (ii) $O(\pd_X)$ off-equilibrium self-energy, combined with the ordinary $O(\pd_X^0)$ resummation equation. (i) and (ii) naturally fit into $D_\text{probe}^{(1)}$ and $D_\text{fluid}^{(1)\text{le}}$ respectively. While the weak coupling studies show polarization dependence in spectral function are generically present from responses to acceleration, shear stress and vorticity for the two cases \cite{Lin:2024svh}, the strongly coupled study has vanishing response to the shear stress in the row of $D_\text{fluid}^{(1)\text{le}}$ in Table.~\ref{tab:summary}.

\section{Conclusion and Outlook}\label{sec:conclusion}

We have studied spectral function of a probe holographic baryon in the fluid-gravity background, with the baryon modeled by a combination of left and right-handed Weyl fermions. The spectral function in a slow-varying background is meaningful when the frequency and momentum of the baryon are much larger than the gradient of the background. We show how to perform gradient expansion for the probe baryon consistently, which includes expansions of the Dirac field and the Dirac operator. The complete gradient correction to the spectral function has been understood as gradient correction to density matrices of the probe baryon and the medium, including $D_\text{probe}^{(1)}$, $D_\text{fluid}^{(1)\text{le}}$ and $D_\text{fluid}^{(1)\text{ss}}$. The corresponding baryon is found to be spin polarized in the presence of fluid acceleration, vorticity and shear stress up to $O(\pd_X)$.

As mentioned in the introduction, modified spectral function found in this study only leads to one contribution to spin polarization. A more relevant quantity is the lesser function of the baryon, which directly connects to what is measured experimentally \cite{Becattini:2013fla}. The lesser function involves solving the Dirac equation in the Schwinger-Keldysh extended fluid-gravity background, which can help reveal more structure in the contribution to spin polarization.

While modeling baryon with bulk Dirac fermion is an obvious first step, the choice is known to violate single-particle spectral sum rule \cite{Gursoy:2011gz}. Real world baryon in vacuum is expect to satisfy the sum rule. It is therefore necessary to adopt more realistic baryon model. The prescription for restoring the sum rule is known in holography \cite{Gursoy:2011gz}. Studies with more realistic baryon models would help identify generic features in spin polarization.

\section*{Acknowledgments}
We thank Umut G\"ursoy and Yan Liu for fruitful discussions. S.L. also thanks Tsinghua Sanya International Mathematics Forum and ECT* for hospitality and providing a stimulus environment during the workshops "Gauge Gravity Duality 2024" and "Holographic Perspectives on Chiral Transport and Spin Dynamics" at the final stage of this work. This work is in part supported by NSFC under Grant Nos 12475148, 12075328 (S.L.) and 12005033 (S-w. L.). Si-wen Li is also supported by the Fundamental Research Funds for the Central Universities under Grant No. 3132025192.

\appendix

\section{The notations}\label{sec:appA}

In this work, the capital letters $L,M,N...$ refer to the indices
of the bulk coordinate as $x^{M}$, the lowercase letters $a,b,c$
refer to the corresponding indices of the coordinate in tangent space.
The gamma matrices $\gamma^{a},\Gamma^{M}$ as the generators of the
Clifford algebra with elfbein $e_{M}^{a}$ are related as,

\begin{equation}
\left\{ \gamma^{a},\gamma^{b}\right\} =2\eta^{ab},\left\{ \Gamma^{M},\Gamma^{M}\right\} =2g^{MN},\Gamma^{M}=e_{a}^{M}\gamma^{a},g_{MN}=e_{M}^{a}\eta_{ab}e_{N}^{a},\text{\tag{A-1}}
\end{equation}
where $g_{MN}$ refers to the metric on a manifold and $\eta^{ab}$
refers to the Minkowskian metric. The five gamma matrices are chosen
as,

\begin{equation}
\gamma^{\mu}=i\left(\begin{array}{cc}
0 & \sigma^{\mu}\\
\bar{\sigma}^{\mu} & 0
\end{array}\right),\gamma^{4}=\gamma=\left(\begin{array}{cc}
1 & 0\\
0 & -1
\end{array}\right),\text{\tag{A-2}}
\end{equation}
with the Pauli matrices $\sigma^{\mu}=\left(1,-\sigma^{i}\right),\bar{\sigma}^{\mu}=\left(1,\sigma^{i}\right),i=1,2,3$.
The spin connection is defined as,
\begin{equation}
\omega_{Mab}=\eta_{cb}\left(e_{a}^{N}\partial_{M}e_{N}^{c}-e_{a}^{N}\Gamma_{MN}^{K}e_{K}^{c}\right),\text{\tag{A-3}}
\end{equation}
where the affine connection $\Gamma_{MN}^{K}$ is given by,
\begin{equation}
\Gamma_{MN}^{K}=\frac{1}{2}g^{KL}\left(\partial_{M}g_{LN}+\partial_{N}g_{ML}-\partial_{L}g_{MN}\right).\text{\tag{A-4}}\label{eq:A4}
\end{equation}
The Greek letters as $\mu,\nu...$ is used to refer to the indices
of coordinate as $x^{\mu}$ at the holographic boundary. The the lowercase
letters $i,j,k,l$ run over the spatial directions of the holographic
boundary i.e. $i,j,k,l=1,2,3$.

\section{The fluid-gravity background}\label{sec_appB}

We collect explicit metric of the the fluid-gravity background \cite{Bhattacharyya:2007vjd} and hydrodynamic relations following from it.
The bulk metric up to first-order in gradient of the fluid velocity and temperature in the local rest frame is given as
\begin{align}
ds^{2}= & ds_{\left(0\right)}^{2}+ds_{\left(1\right)}^{2},\nonumber \\
ds_{\left(0\right)}^{2}= & 2dtdr-r^{2}f\left(r\right)dt^{2}+r^{2}\delta_{ij}dx^{i}dx^{j},\nonumber \\
ds_{\left(1\right)}^{2}= & -2x^{\mu}\partial_{\mu}u_{i}drdx^{i}-2x^{\mu}\partial_{\mu}u_{i}r^{2}\left(1-f\right)dtdx^{i}-\frac{4x^{\mu}\partial_{\mu}b}{r^{2}}dt^{2}\nonumber \\
 & +2r^{2}bF\left(r\right)\sigma_{ij}dx^{i}dx^{j}+\frac{2}{3}r\partial_{i}u_{i}dt^{2}+2r\partial_{0}u_{i}dtdx^{i},\text{\tag{B-1}}\label{eq:B1}
\end{align}
with $x^{\mu}=\left\{ t,x^{i}\right\}$ in the EF coordinates.
Here we have written the metric in the unit of $R=1$ (the radius
of bulk). 
The relevant functions presented in (\ref{eq:B1}) are given as
\begin{align}
\sigma_{ij} & =\partial_{i}u_{j}+\partial_{j}u_{i}-\frac{2}{3}\delta_{ij}\partial^{k}u_{k},\nonumber \\
f\left(r\right) & =1-\frac{1}{b^{4}r^{4}},\ b=\frac{1}{\pi T}=r_{H}^{-1},\nonumber \\
F\left(r\right) & =\int_{br}^{\infty}dx\frac{x^{2}+x+1}{x\left(x+1\right)\left(x^{2}+1\right)}=\frac{1}{4}\left[\ln\left[\frac{\left(1+br\right)^{2}\left(1+b^{2}r^{2}\right)}{b^{4}r^{4}}\right]-2\arctan\left(br\right)+\pi\right].\text{\tag{B-3}}
\end{align}
The following hydrodynamic relations follow from the background 
\begin{equation}
\partial_{0}b=\frac{1}{3}\partial_{i}u^{i},\partial_{i}b=\partial_{0}u_{i}.\text{\tag{B-4}}
\end{equation}
%

\bibliographystyle{unsrt}\bibliography{polarization_fluid.bib}



\end{document}